# Hallmarks of terahertz magnon currents in an antiferromagnetic insulator

Hongsong Qiu[1,2,*], Oliver Franke[2,*], Yuanzhe Tian[3,*], Zdeněk Kašpar[2], Reza Rouzegar[2], Oliver Gueckstock[2], Ji Wu[1], Maguang Zhu[4], Biaobing Jin[5], Yongbing Xu[1], Tom S. Seifert[2], Di Wu[3,†], Piet W. Brouwer[2,†], Tobias Kampfrath[2,†]

[1]State Key Laboratory of Spintronics Devices and Technologies, School of Integrated Circuits, Nanjing University, Suzhou 215163, PR China

[2] Department of Physics, Freie Universität Berlin, Berlin 14195, Germany

[3]National Laboratory of Solid-State Microstructures, Jiangsu Provincial Key Laboratory for Nanotechnology, Collaborative Innovation Center of Advanced Microstructures and Department of Physics, Nanjing University, Nanjing 210093, PR China

[4]School of Integrated Circuits, Nanjing University, Suzhou 215163, PR China

[5]Research Institute of Superconductor Electronics, School of Electronic Science and Engineering, Nanjing University, Nanjing 210023, PR China

[†]Corresponding author.

**Abstract.** The efficient transport of spin angular momentum is expected to play a crucial role in future spintronic devices, which potentially operate at frequencies reaching the terahertz range. Antiferromagnetic insulators exhibit significant potential for facilitating ultrafast pure spin currents by terahertz magnons. Consequently, we here use femtosecond laser pulses to trigger ultrafast spin currents across antiferromagnetic NiO thin films in Py|NiO|Pt stacks, where permalloy (Py) and Pt serve as spin-current source and detector respectively. We find that the spin current pulses traversing NiO reach a velocity up to 40 nm/ps and experience increasing delay and broadening as the NiO thickness is increased. We can consistently explain our observations by ballistic transport of incoherent magnon. Our approach has high potential to characterize terahertz magnon transport in magnetic insulators with any kind of magnetic order.

**Introduction.** The primary objective of spintronics research is to develop computing and memory devices that leverage spin while exhibiting minimal energy dissipation. The pure spin current transported in magnetic insulators in the form of magnons is particularly intriguing for this purpose due to the absence of Joule dissipation [1]. In this respect, antiferromagnetic insulators are an important class of magnetic materials because they may facilitate magnon transport in the terahertz (THz) range. Their characteristics, such as low magnetic damping, enhanced transverse spin correlation near the Néel point [2], and the superfluidity of Bose-Einstein condensates at low temperature [3,4], position AFIs as natural candidates for efficient spin transport.

The mechanism underlying magnon propagation in antiferromagnetic insulators remains a subject of intense debate, particularly concerning whether spins are transported coherently or not [2]. Due to the magnon gap of antiferromagnets, spin currents injected via thermal gradients [5–8] or radio-frequency resonance [9–12] typically dissipate over transport distances of less than 10 nm. Nevertheless, spin transport over distances reaching up to several micrometers was observed through nonlocal spin detection in antiferromagnetic insulators with weak net magnetization [13–16]. This remarkably long-distance spin transport was attributed to the coherent propagation of well-defined modes [17]. Further experimental and theoretical investigation into spin transport in AFIs is crucial to fully harness their potential as active elements in spintronic technologies.

In the THz range, a spin voltage at the interface of a ferromagnetic-metal layer (FM) and an antiferromagnetic-insulator layer (AFI) can be generated through ultrafast optical excitation of the FM [18]. Using THz emission spectroscopy, time-resolved information of spin transport within the AFI can be obtained [19], offering potential significance for the development of ultrafast AFI-based devices. However, it remains an open question whether the far-from-equilibrium THz magnon modes excited in the AFI propagate coherently or incoherently.

**In this work**, we study the ultrafast dynamics of spin currents across antiferromagnetic NiO thin films by using ultrabroadband THz spin-conductance spectroscopy on a Py|NiO|Pt stack. Clear spin-current transients are observed for a NiO thickness up to 30 nm. The ultrafast spin-transport dynamics are well described by a model based on ballistic transport of incoherent magnons. Our findings provide a direct visualization of spin transport mediated by THz short-wavelength magnons in antiferromagnets. Therefore, our results are relevant for fundamental magnon studies and the development of ultrafast and low-dissipation AFI-based spintronic devices.

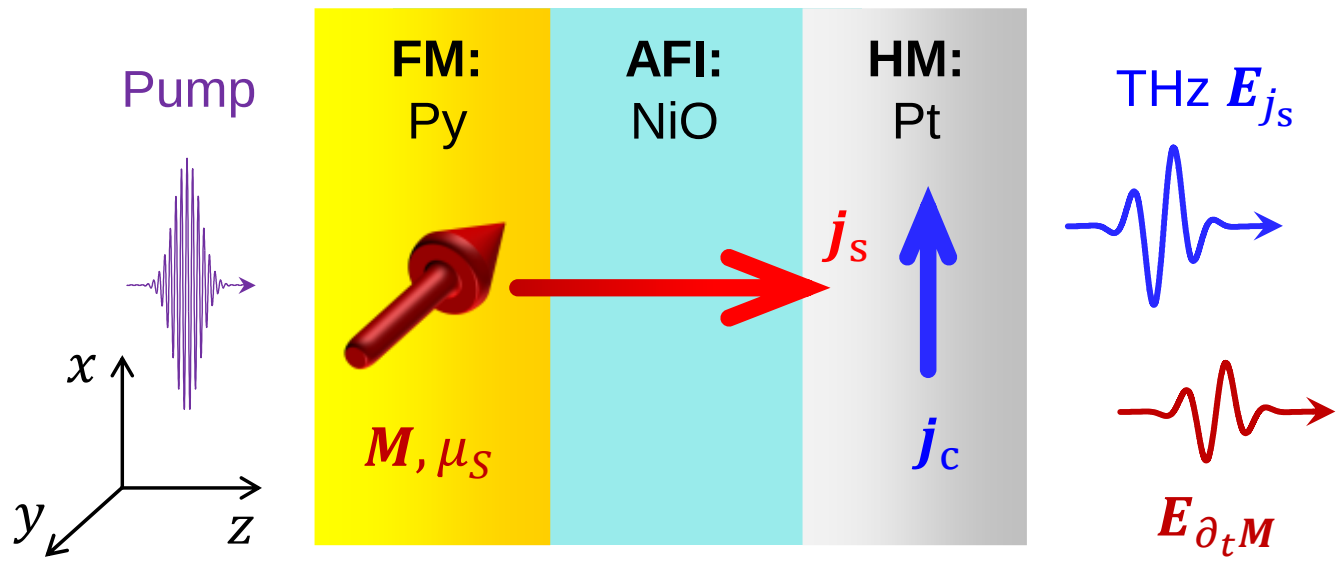

**Figure 1.** Schematic for studying ultrafast spin transport through an antiferromagnetic-insulator layer AFI of NiO in a FM|AFI|HM stack. A femtosecond pump pulse induces a spin voltage $\mu_s(t)$ in the ferromagnetic-metal layer FM of permalloy (Py), which drives a spin current through the AFI. The spin current $j_s(t)$ arriving in the heavy-metal layer HM of Pt is converted into a charge current $j_c(t) \propto j_s(t)$ and monitored via the THz electric field $E_{j_s}(t)$ it emits. Ultrafast quenching of the FM magnetization $\boldsymbol{M}$ leads to the emission of an additional field $E_{\partial_t M}(t)$. An external magnetic field sets the direction of the FM magnetization $\boldsymbol{M}$ along the $y$ axis.

**Concept.** Figure 1 illustrates our THz spin-conductance approach [20] for measuring ultrafast spin transport across a layer AFI made of NiO in a FM|AFI|HM stack, where the ferromagnetic layer FM has magnetization $\boldsymbol{M} = M\boldsymbol{u}_y$, where $\boldsymbol{u}_y$ is the unti vector along the $y$ axis. A femtosecond laser pulse heats the FM transiently and induces a time $(t)$-dependent spin voltage $\mu_\mathrm{s}(t)$ that launches a spin current from FM through AFI into HM. The spin current reaches HM with density $j_\mathrm{s}(t)$ where it is converted into an in-plane charge current $j_\mathrm{c}(t)$ by the inverse spin Hall effect, resulting in a measurable THz electric field $E_{j_\mathrm{s}}(t)$ directly behind the sample. By comparing $E_{j_\mathrm{s}}(t)$ for a sample with an AFI of thickness $d$ and a reference sample FM|HM without AFI ($d = 0$), we can determine the spin conductance $G_\mathrm{AFI}(t)$ of the AFI. Here, $G_\mathrm{AFI}(t)$ represents the spin current that would be obtained for an impulsive spin voltage $\mu_\mathrm{s}(t) \propto \delta(t)$ [20].

We note that, beyond magnon transport across the AFI layer, there are more mechanisms leading to magnetic-field dependent THz-field emission, which need to be considered [21]. In particular, heating by the pump pulse also triggers spin transfer to the crystal lattice and changes the FM magnetization $M$ at a rate $\partial_t M(t) \propto \mu_\mathrm{s}(t)$, resulting in an additional electric field $E_{\partial_t M}(t)$ (Fig.1) [22,23].

**Experimental setup.** We study Al|FM|AFI|HM|sub stacks with Py(3 nm) as FM, Pt(4 nm) as HM and Al as a cap layer on a MgO(500 µm) substrate (sub). For AFI, we use NiO layers with a thickness $d$ ranging from 0 to 30 nm (Supplementary). The samples are excited with linearly polarized femtosecond pump pulses (nominal duration 10 fs, energy 3 nJ, photon energy 1.6 eV) from a Ti:sapphire oscillator (repetition rate 80 MHz). The emitted THz electromagnetic field is detected by electro-optic sampling using a GaP(250 µm) (bandwidth 1-7 THz) or ZnTe(1 mm) crystal (1-5 THz, yet higher sensitivity), resulting in the electro-optic signal $S(t, M)$ (Supplementary). All measurements are performed at room temperature in dry air.

The FM magnetization $\pm M$ is set by an external magnetic field of ±100 mT. We focus on the signal component $S(t) = [S(t, +M) - S(t, -M)]/2$ odd in $M$ because it includes the spin transport indicated in Fig. 1. The even-in-$M$ component of the THz signals is much weaker (Supplementary) and attributed to off-resonant AFI spin pumping in the NiO|Pt regions [24] [25].

**Raw data.** Figure 2(a) displays typical THz electro-optic signals $S(t)$ odd in $M$ from Py|NiO($d$)|Pt for $d = 0$, 3, 8, 18 and 30 nm, as measured with a ZnTe(1 mm) detector. The strong THz signal from the reference sample Pt|Py (i.e., $d = 0$) is rescaled by a factor of 1/20. All signals $S(t)$ exhibit similar early dynamics ($t < 1$ ps). Importantly, the later dynamics ($t > 1$ ps) depends sensitively on $d$ up to 30 nm [Fig. 2(a)]. We confirm that the odd-in-$M$ signal $S(t)$ is independent of the pump polarization and sample azimuth. The associated THz electric field is perpendicular to $\boldsymbol{M}$, consistent with the contributions $E_{j_\mathrm{s}}$ and $E_{\partial_t M}$ indicated in Fig. 1.

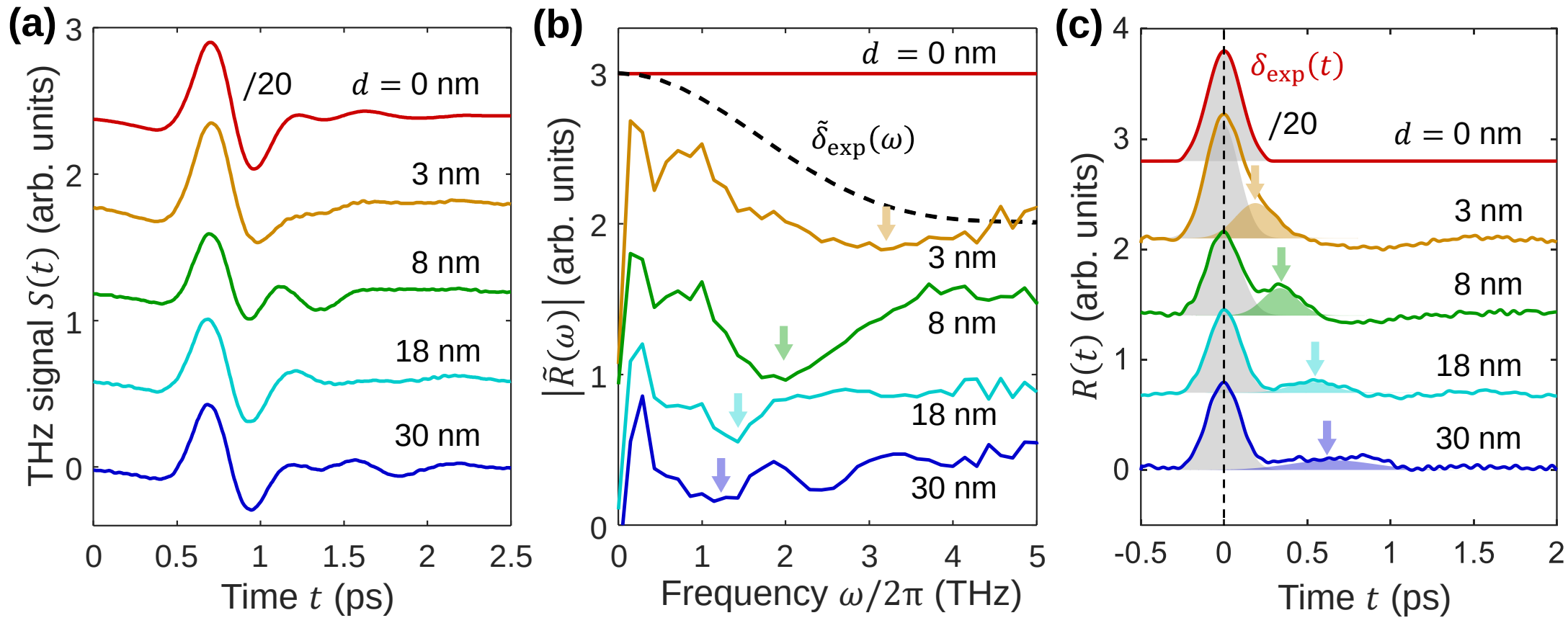

**Figure 2.** THz-emission signals from optically excited Py|NiO($d$)|Pt stacks. (a) Electro-optic signal $S(t)$ for various NiO thicknesses $d$, measured using a 1 mm ZnTe crystal. Note the scaling factor for the reference signal $S_{\text{ref}}(t) = S(t)|_{d=0}$ from Py|Pt. (b) Ratio $\tilde{R}(\omega)$ of the Fourier amplitude spectra of $S(t)$ and $S_{\text{ref}}(t)$ [Eq. (1)]. Arrows indicate spectral dips. The filter function $\tilde{\delta}_{\text{exp}}(\omega)$ (dashed line) represents the finite frequency bandwidth of our setup. (c) The function $(R * \delta_{\text{exp}})(t)$ obtained by inverse Fourier transformation of $\tilde{R}(\omega)\tilde{\delta}_{\text{exp}}(\omega)$ from panel (b). The shaded areas at $t = 0$ (dashed line) and at $t > 0$ (arrows) indicate the two components making up $R(t)$. The function $\delta_{\text{exp}}(t)$ is the inversely Fourier-transformed $\tilde{\delta}_{\text{exp}}$ and represents the impulsive spin voltage. In all panels, curves for different thicknesses are offset horizontally for better visibility.

**Frequency-domain.** Let us, for the moment, assume that the total transient THz electric field $E$ is dominated by the sum $E_{j_s} + E_{\partial_t M}$ (Fig. 1), where both $E_{j_s}$ and $E_{\partial_t M}$ are driven by the transient spin accumulation $\mu_s$ in FM. Note that any features of spin transport ($j_s$) and local magnetization quenching ($\partial_t M$) are strongly blurred in the measured electro-optic signal traces $S = S_{j_s} + S_{\partial_t M}$ of Fig. 2(a). Reasons are the non-instantaneous dynamics $\mu_s(t)$ of the spin accumulation in FM, the conversion of the charge current arising from $j_s$ and $M$ into an electric field, the propagation of $E$ to the detector and its electro-optic detection (Fig. 1) [20,26].

To remove the impact of all of these processes from the measured signal $S = S(t)$, we note that the reference signal $S_{\text{ref}}(t) = S(t)|_{d=0}$ from the reference sample Py|Pt is dominated by spin transport from FM directly into HM and, thus, contains the dynamics $\mu_s(t) \propto j_s(t)|_{d=0}$ of the spin voltage in the most direct way. It shows in Supplementary that, in the frequency domain, the ratio of $S$ and $S_{\text{ref}}$ fulfills

$$\tilde{R}(\omega) = \frac{\tilde{S}(\omega)}{\tilde{S}_{\text{ref}}(\omega)} = \frac{\tilde{G}_{\text{AFI}}(\omega)}{g_{\text{ref}}} + b, \qquad (1)$$

where $\tilde{S}(\omega)$ and $\tilde{S}_{\text{ref}}(\omega)$ is, respectively, the Fourier transformed $S$ and $S_{\text{ref}}$ at frequency $\omega/2\pi$. The $\omega$-independent coefficients $1/g_{\text{ref}}$ and $b$ quantify the respective weight of the $j_s$ and $\partial_t M$ contribution (Fig. 1) to the total THz signal. In Eq. (1), the $\partial_t M$ component is frequency-independent because $\partial_t M(t)$ is directly proportional to the instantaneous driving $\mu_s(t)$. Importantly, the quantity $\tilde{G}_{\mathrm{AFI}}(\omega) = \tilde{j}_s(\omega)/\tilde{\mu}_s(\omega)$ is the spin conductance of layer AFI, and $g_{\mathrm{ref}}$ is the spin conductance of the Py/Pt interface [20].

Fig. 2(b) shows $|\tilde{R}(\omega)|$ for various NiO thicknesses $d$. By definition, we have $|\tilde{R}(\omega)| = 1$ for $d = 0\ \mathrm{nm}$ [Eq. (1)]. Interestingly, for $d > 0\ \mathrm{nm}$, we observe the emergence of a spectral dip, whose width and center frequency decrease as $d$ increases from 3 nm to 30 nm. To summarize, $\tilde{R}(\omega)$ sensitively and systematically changes with the NiO thickness, suggesting a significant contribution of spin transport and, thus, $\tilde{G}_{\mathrm{AFI}}(\omega)$ [Eq. (1)] to the THz signal.

**Time domain.** To better understand the spectra in Fig. 2(b), we Fourier-transform $\tilde{R}(\omega)$ into the time domain. According to Eq. (1), the resulting $R(t)$ is expected to be expressed as

$$R(t) = \frac{G_{\mathrm{AFI}}(t)}{g_{\mathrm{ref}}} + b\delta(t). \tag{2}$$

Here, the time-domain spin conductance $G_{\mathrm{AFI}}(t)$ can be understood as the spin current $j_s(t)$ that arrives in HM following its generation by an impulsive spin accumulation $\mu_s(t) \propto \delta(t)$ in FM. Likewise, the term $b\delta(t)$ reflects the rate of change $\partial_t M(t)$ driven by the same $\mu_s(t) = \delta(t)$.

Before Fourier-transforming the measured $\tilde{R}(\omega)$, we apply a Norton-Beer window function $\tilde{\delta}_{\mathrm{exp}}(\omega)$ to restrict $\tilde{R}(\omega)$ to the usable bandwidth $\omega/2\pi = 0$ to 5 THz [Fig. 2(b)] [27]. In the time domain, $\delta_{\mathrm{exp}}(t)$ is a unipolar peak of width 120 fs and can be considered as an approximation to the ideal $\delta(t)$ within the time resolution of our experiment [Fig. 2(c)] [28]. The filtered Fourier spectra $\tilde{R}(\omega)\tilde{\delta}_{\mathrm{exp}}(\omega)$ of Fig. 2(b) yield the time-domain signals $(R * \delta_{\mathrm{exp}})(t)$ shown in Fig. 2(c). Note that the convolution ($*$) of $R$ with $\delta_{\mathrm{exp}}$ smoothens any features of $R$ that are finer than the width of $\delta_{\mathrm{exp}}$, but otherwise leaves the overall shape of $R$ unchanged.

For all NiO thicknesses $d$, the $R * \delta_{\mathrm{exp}}$ traces exhibit a peak around $t = 0$ whose shape agrees well with the experimental $\delta$-function $\delta_{\mathrm{exp}}$ [Fig. 2(c)]. The initial peak is followed by a delayed feature whose peak position increases monotonically with $d$. This behavior is fully compatible with Eq. (2). To summarize, we can consistently ascribe the $\delta$-like peak of the measured $R * \delta_{\mathrm{exp}}$ at $t = 0$ to local ultrafast demagnetization ($\partial_t M \propto \mu_s = \delta_{\mathrm{exp}}$) and the second, time-delayed peak to spin transport through the NiO layer ($j_s = G_{\mathrm{AFI}} * \mu_s = G_{\mathrm{AFI}} * \delta_{\mathrm{exp}}$).

**THz spin conductance.** As we are primarily interested in ultrafast spin transport through NiO, we extract the spin conductance $G_{\mathrm{AFI}}(t)$ by subtracting the initial $\delta_{\mathrm{exp}}(t)$-peak from the $(R * \delta_{\mathrm{exp}})(t)$ traces in Fig. 2(c) (Eq. ( 2 ) and Supplementary). The resulting $G_{\mathrm{AFM}}(t)$, convoluted with our time resolution $\delta_{\mathrm{exp}}(t)$, are shown in Fig. 3(a) for $d = 3, 8, 18, 30$ nm.

For a quantitative assessment of the impact of $d$ on the spin conductance, we fit a Gaussian to the peak-like $G_{\mathrm{AFI}}(t)$ (Supplementary) and extract the amplitude [Fig. 3(b)], full width at half maximum (FWHM) [Fig. 3(c)] and time delay $t_{\mathrm{pk}}$ [Fig. 3(d)] of $G_{\mathrm{AFI}}(t)$. The amplitude of $G_{\mathrm{AFI}}$ decreases on a length scale of 10 nm [Fig. 3(b)]. The FWHM increases monotonically for $d \geq 8$ nm to 0.36 ps at 30 nm, but exhibits a local maximum at $d = 3$ nm [Fig. 3(c)]. The peak time $t_{\mathrm{pk}}$ increases approximately linearly with $d$ for $d < 10$ nm and slightly sublinearly for $d > 10$ nm [Fig. 3(d)]. Consequently, the resulting signal velocity $d/t_{\mathrm{pk}}$ increases from 22 nm/ps at $d = 3$ nm to 37 nm/ps at 30 nm [Fig. 3(e)].

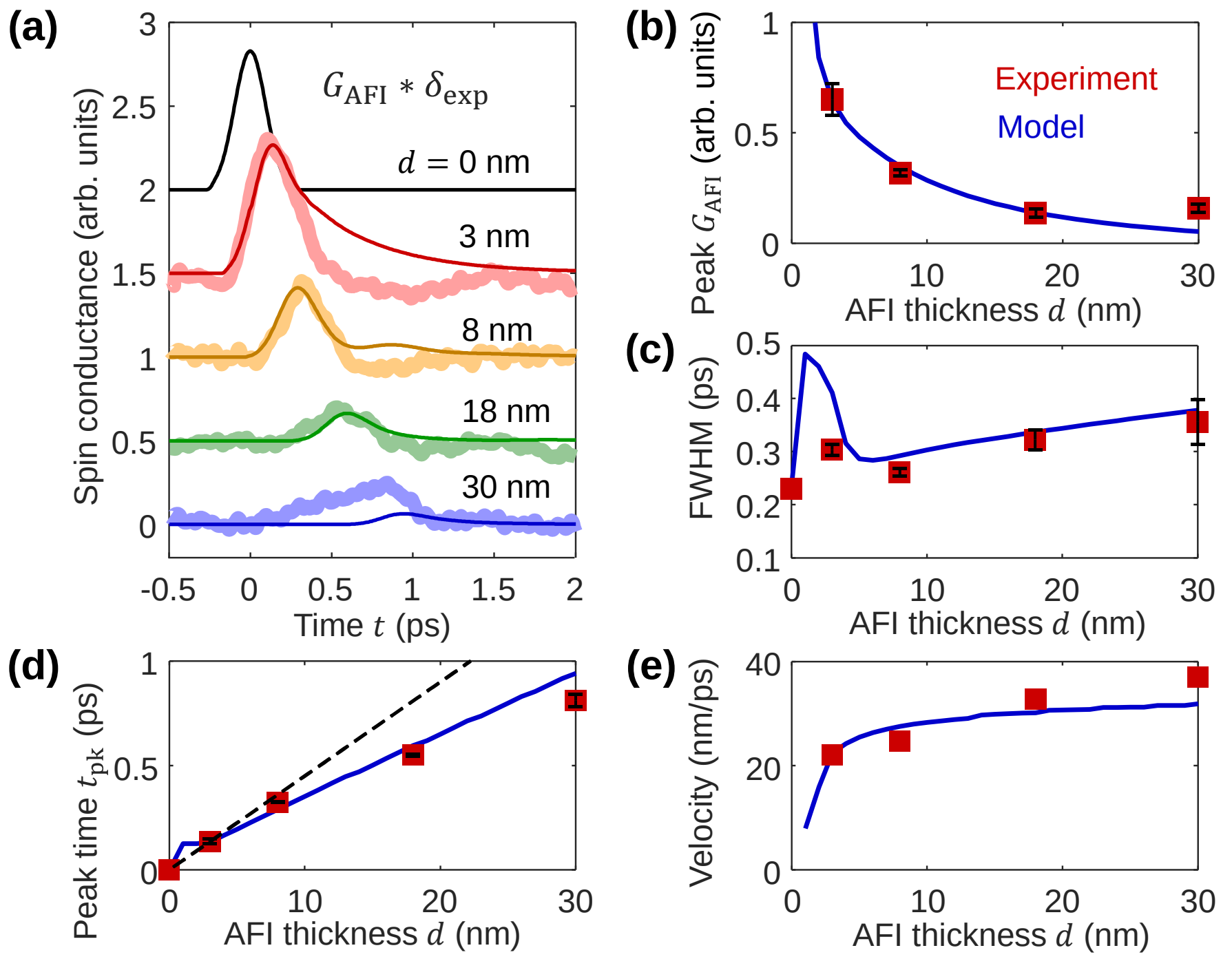

**Figure 3.** (a) Time-domain spin conductance $(G_{\mathrm{AFI}} * \delta_{\mathrm{exp}})(t)$ of the NiO layer for various NiO thicknesses $d$. The convolution with $\delta_{\mathrm{exp}}(t)$ accounts for the experimental time resolution. (b) Extracted amplitude, (c) FWHM, (d) peak time $t_{\mathrm{pk}}$ and (e) mean velocity $d/t_{\mathrm{pk}}$ of the peak-like $(G_{\mathrm{AFI}} * \delta_{\mathrm{exp}})(t)$ as a function of the NiO thickness $d$.

**Interpretation.** Importantly, the velocities of Fig. 3(e) agree with the magnon group velocities of NiO at magnon wavevectors larger than 5% of the Brillouin-zone extent [Fig. 4(a)]. This observation is significant because, as a magnetic insulator, NiO permits spin transport primarily by magnons, tunneling, or by metallic imperfections like pinholes. For comparison, in insulating MgO, transport of spin-polarized electrons through pinholes and by tunneling was found to decrease by more than 1 order of magnitude over just 1 nm of MgO [20], 2 orders of magnitude more rapidly than the observed decrease of $G_{\mathrm{AFI}}$ vs $d$ in NiO [Fig. 3(b)]. We conclude that the measured $G_{\mathrm{AFI}}(t)$ [Fig. 3(a)] can consistently be interpreted as spin conductance that is dominated by magnon transport.

We, thus, suggest that the following scenario dominates the physics underlying our experimental observations. Pump-induced heating of FM induces a change $\Delta T_{\mathrm{e}}$ in the electronic temperature in FM, whereas AFI remains unexcited. The $\Delta T_{\mathrm{e}}$ results in a spin voltage $\mu_s$ in FM that injects spin and, thus, magnons into the AFI through spin-transfer torque. Once the magnons have reached the AFI/HM interface, a fraction of the magnon current enters HM by spin pumping.

We can draw three relevant conclusions on the magnon transport. First, if the spin transport was predominantly diffusive, we would expect the emergence of a diffusive tail in the magnon-current waveforms $G_{AFI}(t)$ [Fig. 3(a)] and a decrease of their average velocity $d/t_{pk}$ because of $t_{pk} \propto d^2$ [Fig. 3(e)]. However, the absence of these features suggests that the magnon transport is rather ballistic-like. Second, specular reflections at the FM/AFI and AFI/HM interfaces are expected to have estimated reflection coefficients above of about 0.6 and 0.8, respectively, and, thus, lead to clear echoes in the $G_{AFI}(t)$ waveforms (Supplementary). In contrast, the observed waveforms [Fig. 3(a)] do not exhibit pronounced echoes, indicating that magnon reflections at the interfaces are rather diffuse. Third, the participating magnons could, in principle, be both coherent and incoherent. If spatially coherent magnons were dominant, they would propagate akin to plane waves along the normal of layer AFI. Given the nearly linear magnon-dispersion relation of NiO within the frequency range covered, the magnon pulses would broaden by less than 10% over a 30 nm propagation distance in NiO (Supplementary). However, the observed broadening of more than 50% [Fig. 3(c)] suggests that the spin-current pulses $G_{AFI}(t)$ [Fig. 3(a)] are predominantly carried by spatially incoherent magnons.

**Model vs experiment.** To test our interpretation quantitatively, we develop the following theoretical description of our experiment based on the linearized Boltzmann transport equation (Supplementary). Spin injection by the pump-induced spin voltage $\mu_s$ in FM is modeled by spin-transfer torque of FM on AFI. The resulting incoherent magnons in AFI start propagating ballistically. Scattering is accounted for by a finite global lifetime $\tau$ of each magnon mode, whereas secondary magnons and, thus, diffusive transport are neglected. Reflection at the FM/AFI and AFI/HM interfaces is modeled as diffuse. Finally, spin transfer from AFI to HM is again modeled by spin pumping and yields the spin-current density $j_s(t)$. The parameters of the simplified 1-band magnon dispersion of NiO are taken from literature, and the spin-mixing conductance of the Py/NiO and NiO/Pt interface is estimated by, respectively, the Sharvin limit and the value of YIG/Pt. A magnon lifetime of $\tau = 0.6$ ps is found to yield the best agreement with the experimental data.

As seen in Fig. 3(a), the calculated spin current $(G_{AFI} * \delta_{exp})(t)$ (solid line) in response to a spin voltage $\delta_{exp}(t)$ agrees well with the measurement (marks) for various NiO thicknesses, except for the largest $d$ of 30 nm. We ascribe this discrepancy to the increasing contribution of diffusive transport that is not accounted for in our model. Our model also well captures the $d$-dependence of the various parameters characterizing $(G_{AFI} * \delta_{exp})(t)$ [Fig. 3(b-e)].

**Discussion.** Our model provides important insights into the ultrafast magnon transport in NiO, as detailed in Supplementary. First, the amplitude decay of the magnon pulse $G_{AFI}(t)$ vs $d$ [Fig. 3(b)] is dominated by the magnon relaxation length $\lambda = v\tau$ of 20 nm, i.e., the product of magnon group velocity and lifetime.

Second, at first glance, one would expect an increase of the FWHM proportional to the propagation length $d$. The reason is that the mutually incoherent magnon wave packets can travel under all angles $\alpha$ relative to the FM/AFI interface normal [Fig. 4(b)]. They cover all velocities along the $z$ axis, from $v$ ($\alpha = 0°$) down to 0 ($\alpha = \pm 90°$), which, respectively, form the leading and trailing edge of $G_{\mathrm{AFI}}(t)$ [29]. However, only magnons with $|\alpha| < \alpha_d$ are relevant because magnons with $|\alpha| > \alpha_d$ need to travel by at least one relaxation length $\lambda = v\tau$ further than magnons with $\alpha = 0°$ and do, thus, not make a significant contribution to the spin current arriving in HM. Therefore, the FWHM of $G_{\mathrm{AFI}}(t)$ scales with $\Delta t = (d/\cos\alpha_d - d)/v$. Because $\alpha_d$ decreases as $d$ increases [see Fig. 4(b)], there is a partial compensation of $\alpha_d$ vs $d$ in $\Delta t$, and the FWHM grows only sublinearly with $d$ (Supplementary).

Third, the nonmonotonic FWHM-vs-$d$ behavior between $d = 0$ and 10 nm [Fig. 3(c)] arises from magnon reflection at the NiO/Pt interface and convolution with our finite time resolution $\delta_{\exp}(t)$ [Fig. 3(a)]. Further, convolution in combination with a wider spread between leading and trailing edge results in a shift of the peak position of $G_{\mathrm{AFI}}(t)$ [Fig. 3(d)] and, thus, increase of the propagation velocity with rising $d$ [Fig. 3(e)] (Supplementary).

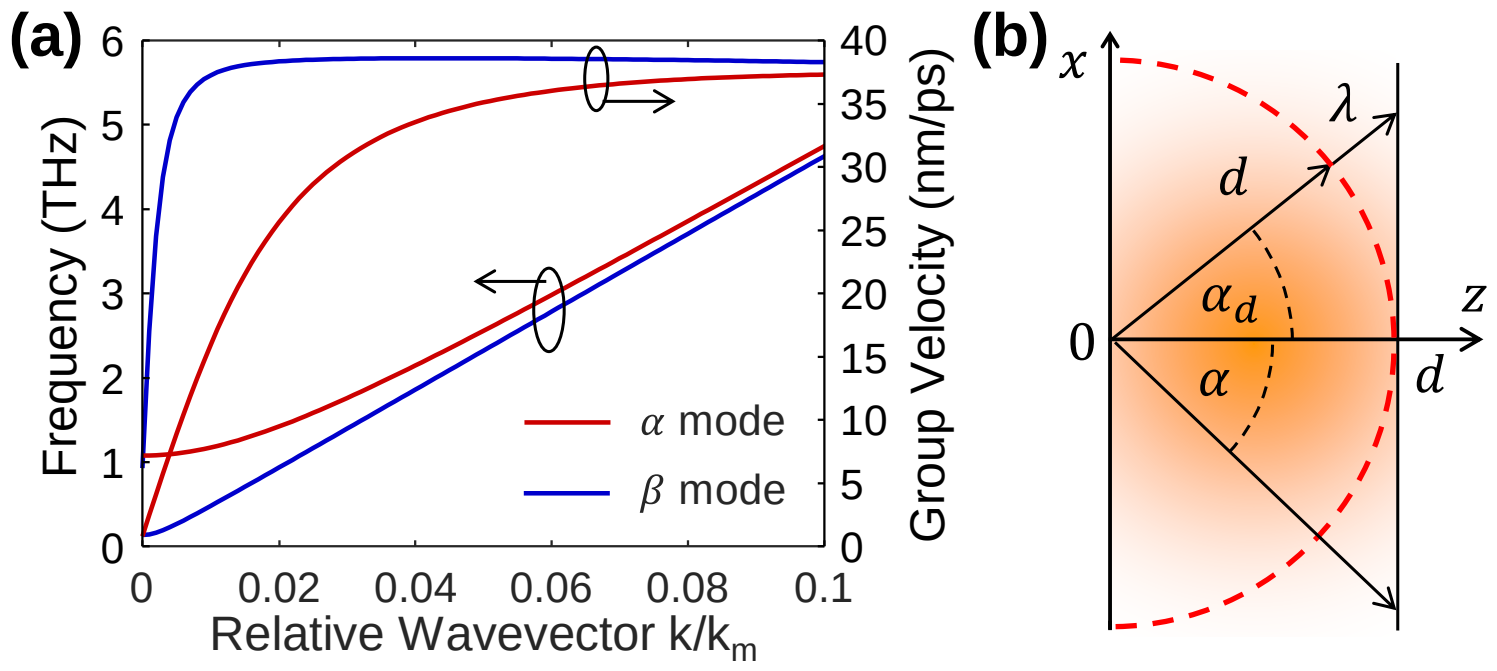

**Figure 4.** Incoherent magnon propagation. (a) Magnon dispersion relation $\Omega_{\boldsymbol{k}}$ of NiO and the group velocity $\partial\Omega_{\boldsymbol{k}}/\partial\boldsymbol{k}$. (b) Schematic of the propagation of the angular momentum of an incoherent magnon pulse originating from the plane at $z = 0$. The angular momentum propagates spherically into the half-space $z > 0$ with group velocity $v$. Such spherical waves are launched throughout the laser-excited region in the $x$-$y$ plane. Propagation directions $\alpha > \alpha_d$ do not contribute substantially to the angular momentum arriving at $z = d$ because their propagation lengths exceed that of the $\alpha = 0°$ direction by at least by at least the magnon relaxation length $\lambda = v\tau$.

Fourth, our model also accounts for a Seebeck-type contribution to $G_{\mathrm{AFI}}(t)$ that is driven by the pump-induced change $\Delta T_{\mathrm{e}}(t)$ in the FM electron temperature. Note that $\Delta T_{\mathrm{e}}(t)$ has very different temporal dynamics than the spin voltage $\mu_{\mathrm{s}}(t)$, which approximately scales with the derivative $\partial_t \Delta T_{\mathrm{e}}(t)$. Therefore, if $G_{\mathrm{AFI}}(t)$ was dominated by the Seebeck contribution, Eq. (1) would be inappropriate, and $\tilde{S}(\omega)/\tilde{S}_{\mathrm{ref}}(\omega)$ would not yield the sum of two unipolar peaks in the time domain (Supplementary). We conclude that a Seebeck-type contribution to the magnon current due to $\Delta T_{\mathrm{e}}(t)$ is negligible.

**Conclusion.** In conclusion, we observe spintronic THz emission from Py|NiO|Pt stacks with antiferromagnetic NiO, whose thickness has a pronounced impact on the emitted THz waveforms. Our data can consistently be explained by ultrafast spin-voltage-driven spin transport through NiO that is predominantly carried by a ballistic flow of incoherent magnons at a high velocity of up to 40 nm/ps and a relaxation time of 0.6 ps. Our work highlights the potential of optically excited FM|AFI|HM stacks as a versatile platform to study THz magnon transport across magnetic insulators with antiferromagnetic order.